\pdfoutput=1
\documentclass[%
 aps, physrev,
 preprint, 
 amsmath,amssymb,
 superscriptaddress, 
 nofootinbib, 
 preprintnumbers, 
]{revtex4-2}
\usepackage{bm}
\usepackage{booktabs}
\usepackage{mathrsfs}

\usepackage[pdftex]{graphicx,color}
\usepackage{color}
\usepackage{comment}

\usepackage[normalem]{ulem}
\usepackage{multirow}

\begin{document}
\title{Lepton flavor violating decay of true muonium:  
 $\boldsymbol{(\mu^+ \mu^-) \to \mu^\pm e^\mp}$}

\author{Ryotaro Minato}
\email[E-mail: ]{minato-ryotaro-vj@ynu.jp}
\affiliation{Department of Physics, Faculty of Engineering Science, 
Yokohama National University, Yokohama 240-8501, Japan}

\author{Akira Sato}
\email[E-mail: ]{sato@phys.sci.osaka-u.ac.jp}
\affiliation{Department of Physics, Graduate School of Science, 
The University of Osaka, 1-1 Machikaneyama, Toyonaka, 560-0043, Osaka, Japan}

\author{Ryosuke Suda}
\email[E-mail: ]{r.suda.813@ms.saitama-u.ac.jp}
\affiliation{Graduate School of Science and Engineering, 
Saitama University, 255 Shimo-Okubo, Sakura-ku, Saitama 338-8570, Japan}

\author{Masato Yamanaka}
\email[E-mail: ]{m.yamanaka.km@cc.it-hiroshima.ac.jp}
\affiliation{Department of Advanced Sciences, 
Faculty of Science and Engineering, 
Hosei University, Tokyo 184-8584, Japan}
\affiliation{Department of Literature, Faculty of Literature, 
Shikoku Gakuin University, Kagawa 765-8505, Japan}
\affiliation{Department of Global Environment Studies, 
Hiroshima Institute of Technology, Hiroshima, 731-5193, Japan}

\preprint{STUPP-25-279}

\date{\today}
\begin{abstract}
We propose a new channel for probing charged lepton flavor violation (CLFV):  the decay of true muonium into a lepton pair of different flavor, $(\mu^+ \mu^-) \to \mu^\pm e^\mp$. 
This purely leptonic two-body decay provides a clean experimental signature in the form of energetic, oppositely charged leptons. It is sensitive not only to photonic dipole interactions but also to four-fermion contact interactions, and is free from hadronic uncertainties in theoretical predictions. 
We evaluate the branching ratios induced by scalar-, vector-, and dipole-type CLFV operators. Our results show that the branching ratio can reach up to $\mathcal{O}(10^{-20})$ within current experimental bounds. 
This decay mode may be discovered as an early-stage physics opportunity of the muon collider program by utilizing the large number of muons produced at its front end.

\end{abstract}
\maketitle
\newpage

\section{Introduction}
\label{sec:Intro}
The observation of charged lepton flavor violation (CLFV) would provide clear evidence of physics beyond the Standard Model (SM)~\cite{Kuno:1999jp, 
Raidal:2008jk, Proceedings:2012ulb, Calibbi:2017uvl, Bernstein:2013hba, 
Davidson:2022jai,Ardu:2022sbt}. 
Extensive efforts are ongoing to search for CLFV processes~\cite{Kuno:2013mha, 
Adamov:2018vin, Bartoszek:2014mya, Baldini:2018nnn, MEGII:2023ltw, 
ARNDT2021165679, Blondel:2013ia, Belle-II:2024sce}, with future experiments aiming for improved sensitivities~\cite{Mu2e-II:2022blh, Maso:2023zjp, 
CGroup:2022tli, Voena:2024vme, Cavoto:2017kub, Kuno:2005mm, 
Belle-II:2022cgf, Banerjee:2022xuw}.
In many scenarios, CLFV processes do not involve the direct production of new particles, but rather probe new physics indirectly through virtual particles in intermediate states.
Therefore, exploring a broad range of CLFV channels is essential to comprehensively uncover the structure of the underlying new physics~\cite{Kuno:1996kv, Farzan:2007us, Koike:2010xr}.

\begin{figure}[bp]
\centering
\includegraphics[width=0.99\textwidth]{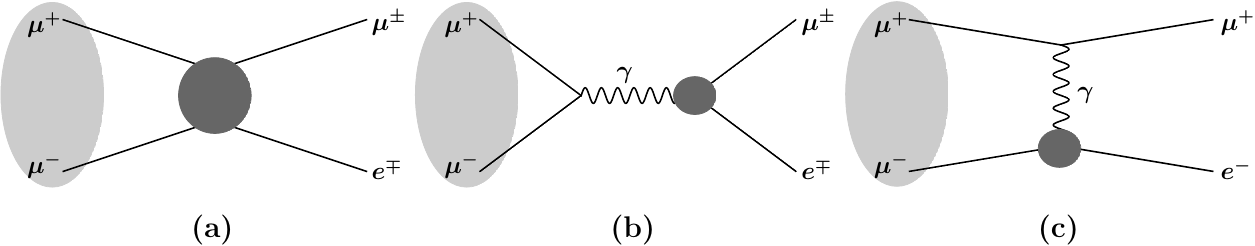}
\caption{CLFV decay of TM for contact interaction (a) and 
for photonic dipole interaction (b,c). Black disks indicate 
the effective CLFV interactions. }
\label{fig:diagram}
\end{figure}

In this study, we focus on true muonium (hereafter abbreviated as TM in the text 
and $(\mu^+ \mu^-)$ in the reaction equations),  
also known as dimuonium, a bound state of  
a negatively charged muon
and a positively charged muon. 
Specifically, we consider the two-body decay
\begin{equation}
\begin{split}
	(\mu^+ \mu^-) \to \mu^\pm e^\mp,
\label{eq:TMCLFVdecay}
\end{split}
\end{equation}
as illustrated in Fig.\ref{fig:diagram}.
This process offers a clean experimental signature: two energetic, oppositely charged leptons in the final state, allowing for full kinematic reconstruction.
Moreover, being a purely leptonic process, it is free from hadronic uncertainties, enabling precise comparison between theoretical predictions and experimental measurements.
The discovery of the TM CLFV decay, when combined with the analysis of relevant 
CLFV processes, e.g., $\mu^+ \mu^- \to \mu^\pm e^\mp$ at collider facilities, 
$\mu \to e \gamma$, etc., sheds light on the ultra-violet structure responsible for CLFV 
interactions. 
Investigating the CLFV decay of TM in a non-vacuum environment or under external magnetic fields may provide access to information that cannot be obtained from collider-based processes such as $\mu^+ e^- \to \mu^+ \mu^-$. For instance, previous studies on analogous bound-state systems have suggested that examining the magnetic-field dependence of the muonium–antimuonium conversion rate can offer insights into the new physics mechanisms responsible for such processes~\cite{Horikawa:1995ae, Hou:1995np}.

TM is a purely quantum electrodynamic system whose existence was theoretically predicted long ago. However, it has not yet been directly observed in experiments. Several experimental groups have initiated or proposed searches for TM. The Heavy Photon Search (HPS) Collaboration~\cite{Banburski:2012tk} is advancing a fixed-target experiment to search for TM, while the DIRAC experiment at CERN has discussed its potential observation in future upgrade phases~\cite{Chliapnikov:1987121}. Recent theoretical studies have also indicated that TM could be produced and detected via ultra-relativistic heavy-ion collisions at the Large Hadron Collider (LHC)~\cite{Bertulani:2023nch}. In addition, exploratory strategies have been proposed involving Belle II and LHCb~\cite{Gargiulo:2025pmu, CidVidal:2019qub}.
With continued experimental progress, TM may be observed within the coming decades, opening new possibilities for precision spectroscopy and new physics searches.
Future high-luminosity  $e^+e^-$ or $\mu^+\mu^-$ colliders may provide environments for copious TM
production~\cite{Fox:2021mdn}. Furthermore, alternative approaches to generate and detect TM by combining low-energy positive and negative muons without relying on beam collisions are also under consideration~\cite{Itahashi:2015fra,Itahashi:2017gpe}.

In this paper, assuming the availability of abundant TM production, we explore the prospects for detecting its CLFV decays.
%
%
The paper is organized as follows. 
In Sec.~\ref{Sec:Int}, we present the CLFV decay rate formula for TM, 
derived from the most general effective Lagrangian.
In Sec.~\ref{Sec:numerical}, first, we briefly review the current bounds on relevant 
CLFV parameters and numerically demonstrate the branching ratio using these bounds 
in the single-operator dominance scenario. Next, we extend our analysis to cases with 
two dominant operators and discuss the experimental signature and potential backgrounds.
Finally, Sec.~\ref{Sec:Summary} provides a summary and outlook.

\section{Interaction, Reaction rate}
\label{Sec:Int}

To evaluate the branching ratio of the CLFV decay $(\mu^+\mu^-) \to \mu^\pm e^\mp$, we work within the framework of effective field theory. 
The effective Lagrangian governing the CLFV decay of TM is given by
\begin{equation}
\begin{split}
    \mathcal{L}
    &=
    - \frac{1}{\Lambda^2} \Bigl[
    g_{LL}^S (\overline{{\psi_e}} P_L \psi_\mu)(\overline{\psi_\mu} P_L\psi_\mu) 
    + g_{RR}^S (\overline{\psi_e} P_R \psi_\mu)(\overline{\psi_\mu} P_R \psi_\mu)  \\
    &~~ 
    + g_{LL}^V (\overline{\psi_e} \gamma^\mu P_L \psi_\mu)(\overline{\psi_\mu} \gamma_\mu P_L \psi_\mu)        
    + g_{RR}^V (\overline{\psi_e} \gamma^\mu  P_R \psi_\mu)(\overline{\psi_\mu} \gamma_\mu P_R \psi_\mu)  \\
    &~~         
    + g_{LR}^V (\overline{\psi_e} \gamma^\mu P_L \psi_\mu)(\overline{\psi_\mu} \gamma_\mu P_R \psi_\mu)
    + g_{RL}^V (\overline{\psi_e} \gamma^\mu P_R \psi_\mu) (\overline{\psi_\mu} \gamma_\mu P_L \psi_\mu)  \\
    &~~ 
    + v A_L \overline{\psi_e} \sigma^{\mu\nu} P_L \psi_\mu F_{\mu\nu} 
    + v A_R \overline{\psi_e} \sigma^{\mu\nu} P_R \psi_\mu F_{\mu\nu} \\
    &~~
    + \text{H.c.}
    \Bigr],
\label{eq:intL}
\end{split}
\end{equation}
where $\sigma^{\mu\nu} = (i/2)[\gamma^\mu,\gamma^\nu]$, $P_L=(1 - \gamma_5)/2$, 
$P_R=(1 + \gamma_5)/2$, $v$ is the vacuum expectation value of Higgs field, and 
$\Lambda$ denotes the energy scale of physics behind the TM CLFV decay.

The branching ratio of TM CLFV decay is given by 
\begin{equation}
\begin{split}
    \text{BR} \left( (\mu^+ \mu^-) \to \mu^\pm e^\mp \right) 
    = 
    \tau_{\text{TM}} \cdot \Gamma \left( \left( \mu^+ \mu^- \right) \to \mu^\pm e^\mp \right), 
\label{eq:BR}
\end{split}
\end{equation}
where $\tau_{\text{TM}}$ represents the lifetime of TM. Both channels 
$(\mu^+ \mu^-) \to \mu^+ e^-$ and $(\mu^+ \mu^-) \to \mu^- e^+$ are included
in the decay rate $\Gamma$.  
In this work, taking unpolarized muons for the 1S initial state, we assume the ratio 
of singlet to triplet is 1:3.
The singlet decays via $2\gamma$ with
$\tau_{\rm TM}(2\gamma)=0.602~{\rm ps}$, and the triplet via $e^+e^-$ with
$\tau_{\rm TM}(e^+e^-)=1.81~{\rm ps}$~\cite{Bilenky:1969zd, TMlifetime2}. 
Weighting by 1:3 yields $\tau_{\text{TM}} = 1.51\, \text{ps} 
= 2.29 \times 10^{12} \, \text{GeV}^{-1}$. 
The decay rate is expressed as
\begin{equation}
\begin{split}
	\Gamma \left( \left( \mu^+ \mu^- \right) \to \mu^\pm e^\mp \right) 
	= 
	 \sigma v_{\text{rel}} |\psi_{\mu}|^2 , 
\label{eq:DecayRate}
\end{split}
\end{equation}
where $v_{\text{rel}}$ is the relative velocity of the initial-state muons,
$|\psi_{\mu}|^2$ represents the overlap of wave functions of the initial muons, and $\sigma$ is 
the cross section of $\mu^+ \mu^- \to \mu^\pm e^\mp$.
We consider the CLFV decay of TM in vacuum, in the absence of both electric and magnetic fields.
Under this environment, the spatial size of TM is approximately the same as 
that of muon, and is estimated as $\mathcal{O} \left( \bigl(\alpha \mu_{\text{TM}} 
 \bigr)^{-1} \right)$ with the analogy of the dynamics of positronium. 
Here $\alpha$ is the fine-structure constant and $\mu_{\text{TM}} = m_\mu/2$ is the reduced mass of TM.
Neglecting the spatial momenta of the initial muons, their 1S wave functions are given by 
\begin{equation}
\begin{split}
    \psi_{\mu}^{1S}(r)=\frac{1}{\sqrt{\pi}}(\alpha \mu_{\text{TM}})^{3/2}
e^{-\alpha \mu_{\text{TM}} r},
\label{eq:WaveFunc}
\end{split}
\end{equation} 
where $r$ is the relative coordinate of the initial muons.

Averaging over initial spins (unpolarized muons) and neglecting the electron mass, 
$\sigma v_{\text{rel}}$ in the nonrelativistic limit of the initial state is calculated to be
\begin{equation}
\begin{split}
    \sigma v_{\text{rel}} = \frac{6}{128 \pi} \frac{m_\mu^2}{\Lambda^4} 
    \left( G^S + G^V + G^D + G^{SV} + G^{SD} + G^{VD} \right), 
\label{eq:XSection}
\end{split}
\end{equation}
where 
\begin{align}
  G^S &\equiv \frac{15}{16}\bigl(\lvert g_{LL}^S\rvert^2 + \lvert g_{RR}^S\rvert^2\bigr), \\[1ex]
  G^V &\equiv 15\bigl(\lvert g_{LL}^V\rvert^2 + \lvert g_{RR}^V\rvert^2\bigr)
         + \frac{51}{4}\bigl(\lvert g_{LR}^V\rvert^2 + \lvert g_{RL}^V\rvert^2\bigr)
         + 6\,\text{Re}\bigl[g_{LL}^V g_{LR}^{V*} + g_{RR}^V g_{RL}^{V*}\bigr], \\[1ex]
  G^D &\equiv \frac{e^2 v^2}{m_\mu^2}\,81\bigl(\lvert A_L\rvert^2 + \lvert A_R\rvert^2\bigr), \\[1ex]
  G^{SV} &\equiv 2\,\text{Re}\Bigl[
             -3\bigl(g_{LL}^S g_{RR}^{V*} + g_{RR}^S g_{LL}^{V*}\bigr)
             + \frac{3}{4}\bigl(g_{LL}^S g_{RL}^{V*} + g_{RR}^S g_{LR}^{V*}\bigr)
           \Bigr], \\[1ex]
  G^{SD} &\equiv \frac{e v}{m_\mu}\,2\,\text{Re}\Bigl[
             \frac{27}{4}\bigl(g_{LL}^S A_L^* + g_{RR}^S A_R^*\bigr)
           \Bigr], \\[1ex]
  G^{VD} &\equiv \frac{e v}{m_\mu}\,2\,\text{Re}\Bigl[
             -27\bigl(g_{LL}^V A_R^* + g_{RR}^V A_L^*\bigr)
             - \frac{27}{4}\bigl(g_{LR}^V A_R^* + g_{RL}^V A_L^*\bigr)
           \Bigr].
\end{align}
The factor $ev/m_\mu \simeq 710$ enhances the reaction rate of the photonic 
interaction channels because the suppression of the ultra violet scale is partially 
cancelled by $v/\Lambda$ for the photonic dipole operator.

To illustrate the parametric dependence of the branching ratio, 
$\text{BR} \equiv \text{BR} \left( (\mu^+ \mu^-) \to \mu^\pm e^\mp \right)$, 
we evaluate it under the single-operator dominance assumption.
For scalar-type dominance model ($g_{LL(RR)}^S \neq 0$ and $\text{others}=0$), 
\begin{equation}
\begin{split}
    \text{BR} = 
    6.54\times 10^{-15} \left( \frac{1\,[\mathrm{TeV}]}{\Lambda} \right)^4 
    \left[ \left| g_{LL}^S \right|^2 + \left| g_{RR}^S \right|^2 \right]. 
\label{eq:BR-S}
\end{split}
\end{equation}
For vector-type dominance model ($g_{LL(RR,LR,RL)}^V \neq 0$ and $\text{others}=0$) ,
\begin{equation}
\begin{split}
    \text{BR} 
    &=
    1.05 \times 10^{-13} \left( \frac{1\,[\mathrm{TeV}]}{\Lambda} \right)^4 
    \Bigl[ 
    \left( \left| g_{LL}^V \right|^2 + \left| g_{RR}^V \right|^2 \right)
    \\& \hspace{-5mm}
    + \dfrac{17}{20} \left( \left| g_{LR}^V \right|^2 + \left| g_{RL}^V \right|^2 \right) 
    + \dfrac{2}{5} \text{Re}\left[ g_{LL}^V g_{LR}^{V*} + g_{RR}^V g_{RL}^{V*} \right] 
    \Bigr], 
\label{eq:BR-V}
\end{split}
\end{equation}
and for dipole-type dominance model ($A_{L(R)} \neq 0$ and $\text{others}=0$), 
\begin{equation}
\begin{split}
    \text{BR} 
    &=
    2.81 \times 10^{-7} \left( \frac{1\,[\mathrm{TeV}]}{\Lambda} \right)^4 
    \left( \left| A_L \right|^2 + \left| A_R \right|^2 \right). 
\label{eq:BR-D}
\end{split}
\end{equation}

\section{Numerical result} 
\label{Sec:numerical}

In this section, we present numerical estimates of the branching ratio 
$\text{BR}((\mu^+\mu^-) \to \mu^\pm e^\mp)$ for representative CLFV scenarios, 
and compare them with existing experimental constraints.

We begin by considering three benchmark models, each dominated by a single type of operator. 
The four-Fermi effective operators introduced in Eq.~\eqref{eq:intL} are generated 
by integrating out heavy degrees of freedom. 
\begin{description}
\item[scalar] A scalar CLFV mediator that universally couples to 
left- and right-handed matter fields. We adopt the single-operator 
dominance model with a universal coupling for chirality; 
$g^S \equiv g_{LL}^S = g_{RR}^S \neq 0$ and omit all other couplings 
in Eq.~(\ref{eq:intL}).
\item[vector-uni.] A vector mediator that universally couples to left- 
and right-handed matter fields, 
where the generated effective operator is 
$(\bar{\psi}_\mu \gamma_\mu \psi_\mu) (\bar{\psi}_\mu \gamma^\mu \psi_e)$. 
Similarly with the scalar case above, the single-operator dominance model is 
adopted, i.e., $g^V \equiv g_{LL}^V = g_{RR}^V = g_{LR}^V = g_{RL}^V \neq 0$. 
\item[vector-L] A vector mediator couples to left-handed 
matter fields only, 
where the generated effective operator is 
$(\bar{\psi}_\mu \gamma_\mu P_{L} \psi_\mu) 
(\bar{\psi}_\mu \gamma^\mu P_{L} \psi_e)$.
\end{description}

\begin{figure}[bp]
\centering
\includegraphics[width=0.55\textwidth]{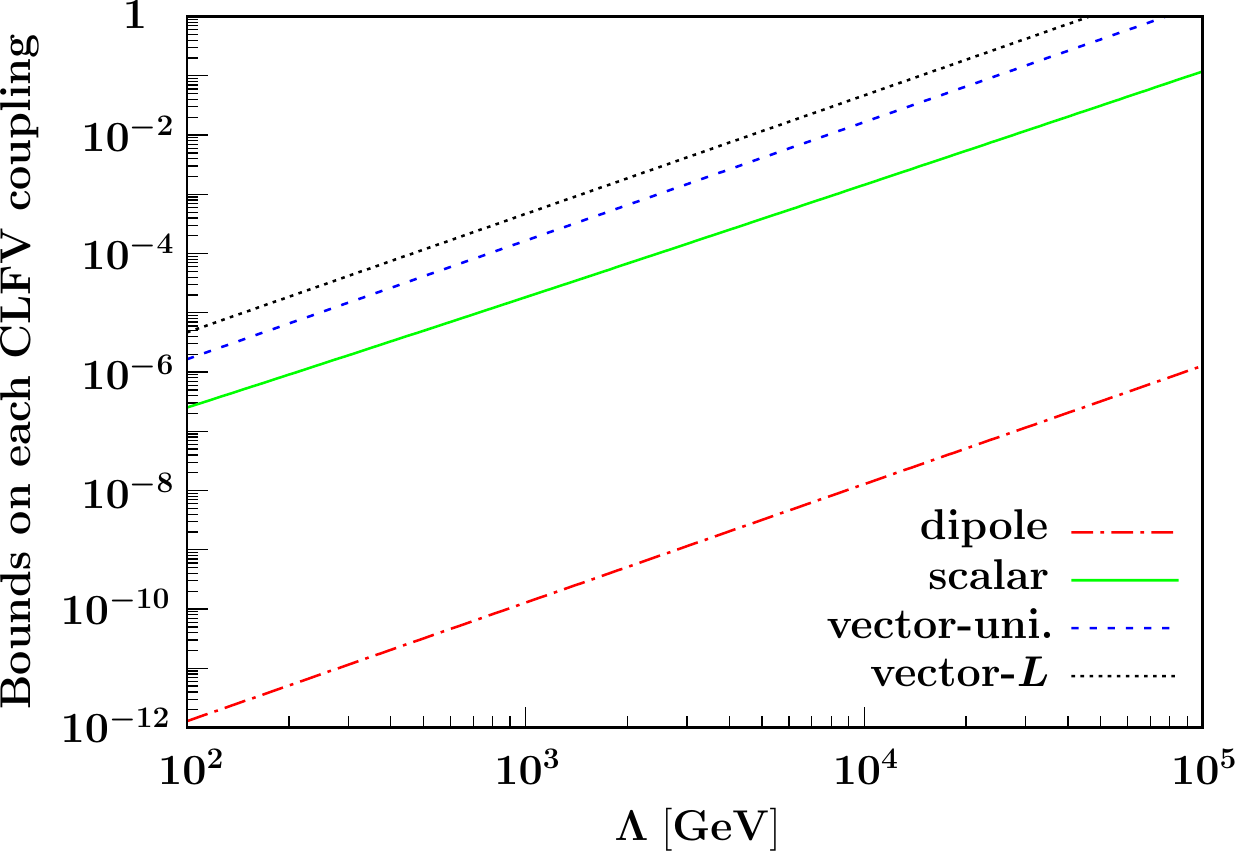}
\caption{Current upper bounds on the effective couplings for each type 
of CLFV operator derived from the MEG II limit on $\mu^+ \to e^+ \gamma$. 
Single-operator dominance is assumed in each case. 
We take $\Lambda$ to be equal to the CLFV mediator mass.}
\label{fig:bound}
\end{figure}

Each of these CLFV operators also induces the radiative decay $\mu^+ \to e^+ \gamma$ via 
loop diagrams, with rates that depend on the distinct Dirac-Lorentz structure of the 
operator~\cite{Lavoura:2003xp}. The most stringent experimental constraint on CLFV comes 
from the MEG II experiment, which has set the upper limit 
$\text{BR} (\mu^+ \to e^+ \gamma) < 3.1 \times 10^{-13}$ at 90\% confidence 
level~\cite{MEGII:2023ltw}.  
Using this experimental upper bound, we derive constraints on the effective CLFV couplings. 
For each scenario, we assume a representative new physics scale of 
$\Lambda = M_\phi = 1$~TeV ($M_\phi$ represents the mass of CLFV mediator) and obtain 
the following bounds:
\begin{equation}
\begin{split}
    \left| g_{LL}^S \right|^2+\left| g_{RR}^S \right|^2 <& \ 3.35\times10^{-10} ~~~ \left( \text{scalar} \right), 
    \\
    \left| g^V \right|^2 <& \ 2.72 \times 10^{-8} ~~~ \left( \text{vector-uni.} \right), 
    \\
    \left| g_{LL}^V \right|^2 <& \ 2.18 \times 10^{-7} ~~~ \left( \text{vector-}L \right), 
    \\
    \left| A_L \right|^2+\left| A_R \right|^2 <& \ 1.64\times10^{-20} ~~~ ( \text{dipole} ).  
\label{Eq:upper}
\end{split}
\end{equation}
Figure~\ref{fig:bound} shows the current bounds on the effective couplings as a function of the new physics scale $\Lambda$. We take $\Lambda$ to be equal to the CLFV mediator mass.

\begin{figure}[tbp]
\centering
\includegraphics[width=0.55\textwidth]{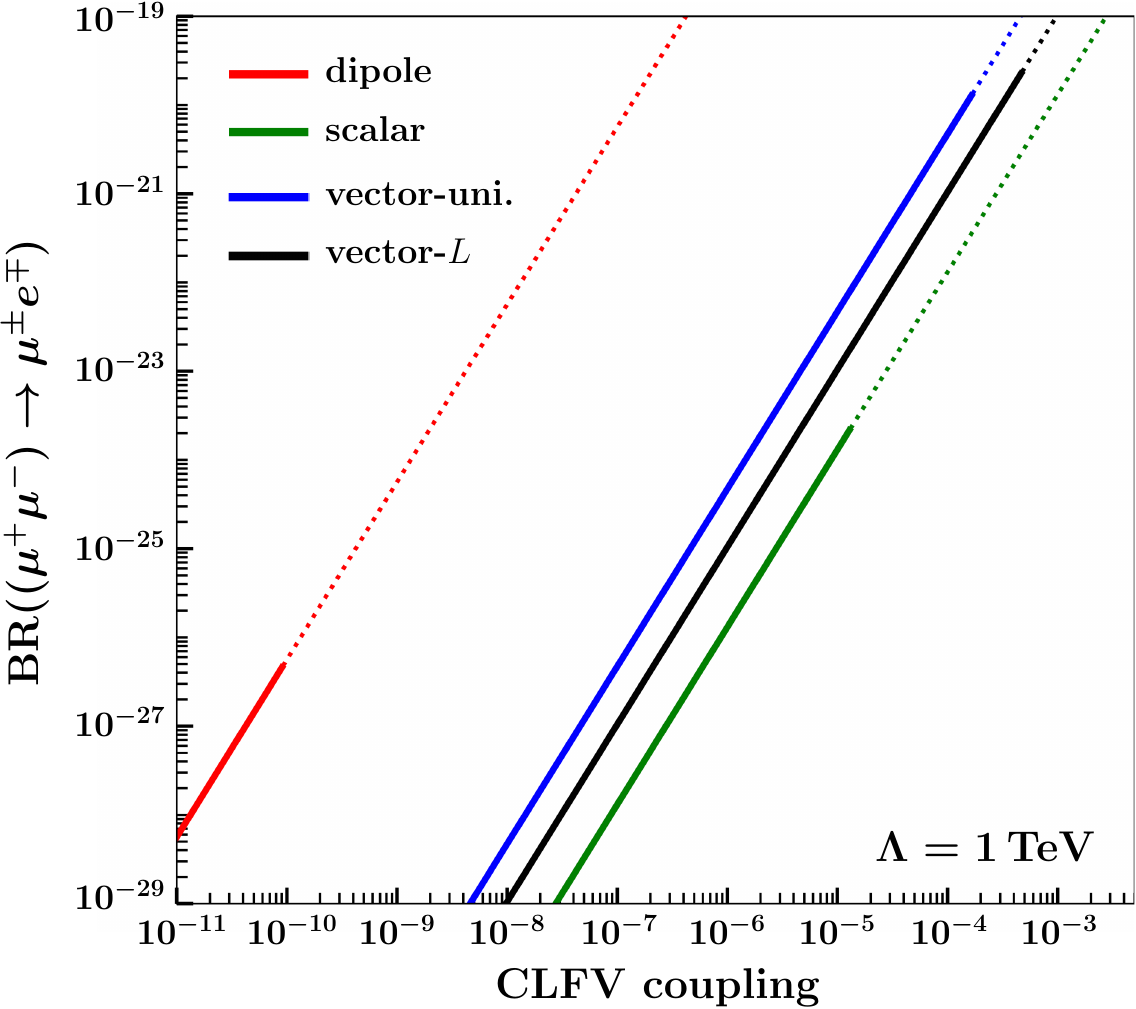}
\caption{Branching ratio $\text{BR}\left( (\mu^+ \mu^-) 
\to \mu^\pm e^\mp \right)$ as a functions of CLFV coupling for 
scalar- (green), vector-uni.- (blue), vector-$L$- (black), 
and dipole- (red) operators. 
Here the new physics scale is taken to be $\Lambda = 1$\,TeV. 
The dashed segments indicate parameter regions excluded by the MEG II constraint on $\mu \to e \gamma$. 
Single-operator dominance is assumed in all cases.}
\label{fig:BR1}
\end{figure}

To assess experimental prospects, we calculate the branching ratio as a function of each 
CLFV coupling under the single-operator dominance assumption. The results are shown in 
Fig.~\ref{fig:BR1}, with $\Lambda = 1$~TeV. The dashed regions indicate parameter ranges 
already excluded by $\mu \to e \gamma$ constraints.
Among the four scenarios, the dipole-type operator yields the most suppressed branching ratio 
due to a more stringent bound compared to other operators. 
This is because the dipole operator contributes to $\mu \to e \gamma$ at tree level, whereas 
scalar and vector operators induce the decay only via loop processes. 
Although we have fixed $\Lambda = 1\,\text{TeV}$ for simplicity, the physical constraints on the couplings depend on the combination $(\text{effective coupling})^2/\Lambda^2$, as is typical for four-Fermi-type interactions.
Since the branching ratio scales as $\Lambda^{-4}$, increasing $\Lambda$ by one order of magnitude suppresses the branching ratio by four orders of magnitude.
As a practical example, we estimate the number of TMs required to detect an event in the vector-$L$ 
scenario. At the maximal allowed $g_{LL}^V$, one needs
\begin{equation}
N_\text{TM} = \mathrm{BR}^{-1} \approx 5.2 \times 10^{19}. 
\end{equation}
At the muon production point of the muon collider currently under consideration, approximately $3 \times 10^{15}$~$\mu^{-}$/s and $2 \times 10^{15}$~$\mu^{+}$/s are expected to be produced~\cite{InternationalMuonCollider:2025sys}. If an efficient method for generating TM using these abundant muons can be developed, there is a possibility that the CLFV decay of TM proposed in this study could be discovered during the early stage of the muon collider program, prior to the construction of ionization cooling and muon acceleration systems.

\begin{figure}[tbp]
  \centering
  \begin{minipage}[t]{0.45\columnwidth}
    \centering
    \includegraphics[width=\columnwidth]{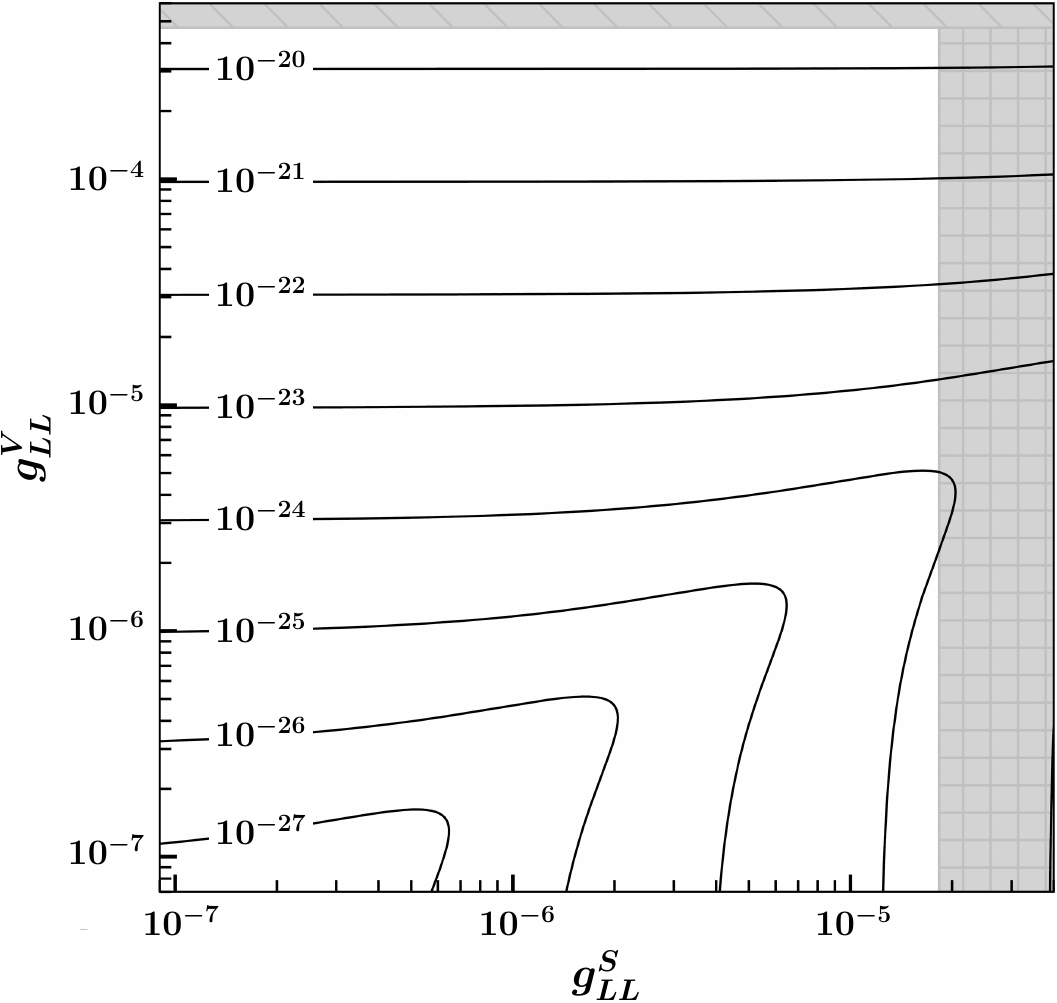}
    {(a) $g_{LL}^S$ -- $g_{LL}^{V}$ plane\label{fig:contour_a}}
  \end{minipage}\quad
  \begin{minipage}[t]{0.45\columnwidth}
    \centering
    \includegraphics[width=\columnwidth]{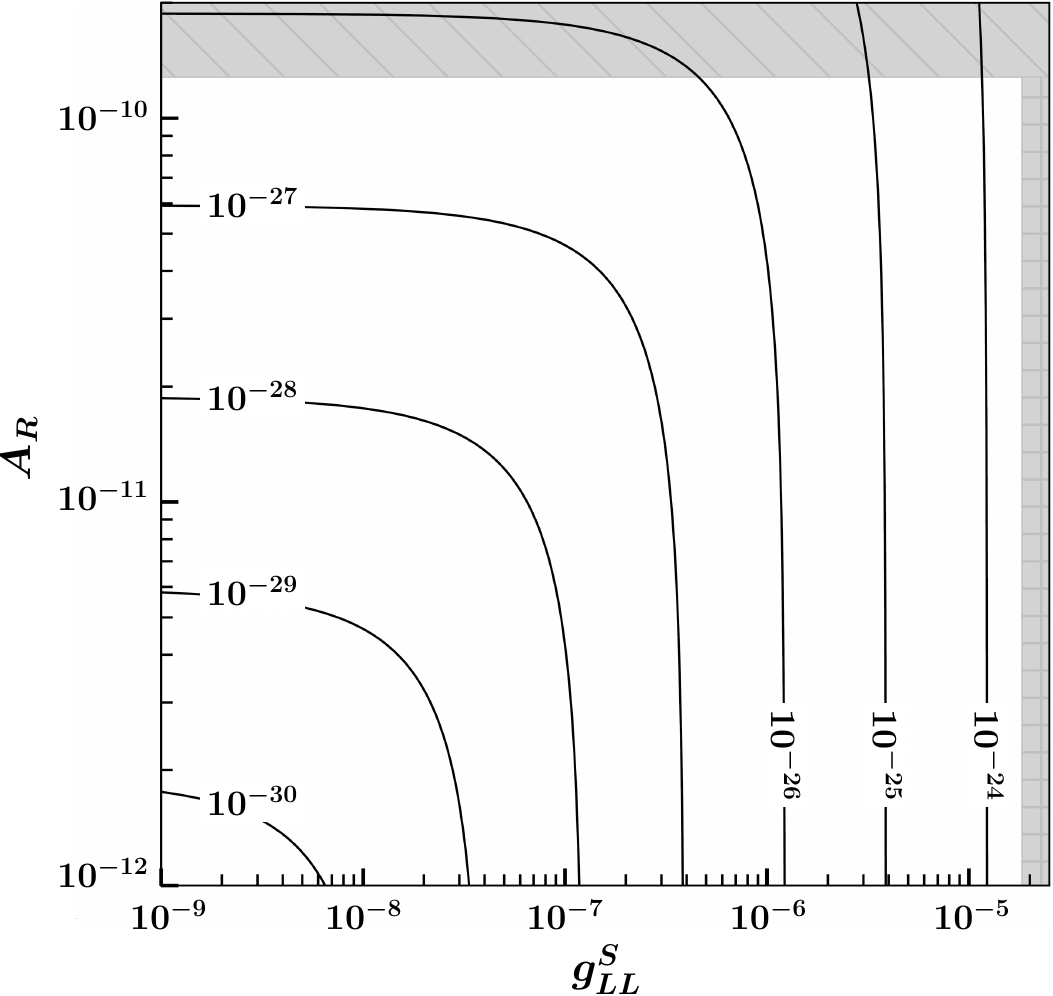}
    {(b) $g_{LL}^S$ -- $A_R^{}$ plane\label{fig:contour_b}}
  \end{minipage}
  \\[5mm]
  \begin{minipage}[t]{0.45\columnwidth}
    \centering
    \includegraphics[width=\columnwidth]{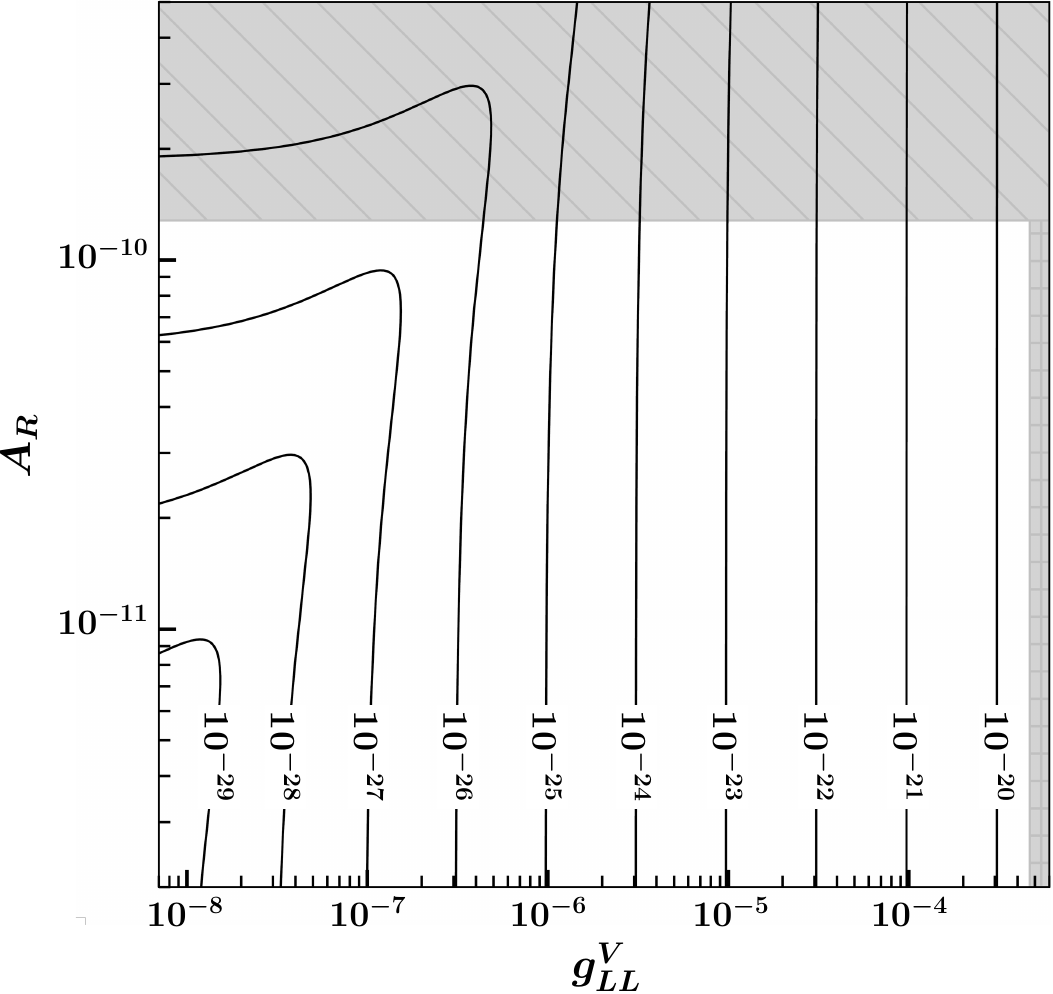}
    {(c) $g_{LL}^V$ -- $A_R^{}$ plane\label{fig:contour_c}}
  \end{minipage}
    \caption{
    Contours of constant branching ratio $\text{BR}((\mu^+\mu^-) \to \mu^\pm e^\mp)$ 
    in the parameter planes of (a) $g_{LL}^S$  --\ $g_{LL}^{V}$, (b) $g_{LL}^S$ 
    --\ $A_R^{}$, and (c) $g_{LL}^V$ --\ $A_R^{}$.
    In each panel, only the two couplings shown on the axes are nonzero and assumed to be real.
    Shaded regions are excluded by the MEG II bound on $\mu \to e \gamma$.
    }
  \label{fig:contour}
\end{figure}

We now generalize the analysis by allowing for multiple operator contributions 
and interference. Figure~\ref{fig:contour} presents contours of constant branching 
ratio in three representative coupling planes:
in the $g_{LL}^S$ -- $g_{LL}^{V}$ plane (Fig.~\ref{fig:contour} (a)), 
in the $g_{LL}^S$ -- $A_R^{}$ plane (Fig.~\ref{fig:contour} (b)), and 
in the $g_{LL}^V$ -- $A_R^{}$ plane (Fig.~\ref{fig:contour} (c)). 
In each case, only the couplings shown are nonzero. These plots allow for 
identification of parameter regions compatible with a hypothetical observation of 
$(\mu^+\mu^-) \to \mu^\pm e^\mp$ and the MEG II limit on $\mu \to e \gamma$.

By including such interference contributions and comparing them with observables 
from other CLFV processes, one can more accurately determine the relevant CLFV 
parameters and the underlying new physics structures. 
For example, consider a scenario in which a CLFV process such as 
$\mu^+e^- \to \mu^+\mu^-$ is observed at a collider, while other muon CLFV processes 
(e.g., $\mu \to e\gamma$, $\mu \text{\,-\,} e$ conversion) remain undetected. 
In this case, it becomes important to investigate the individual contributions of 
the scalar and vector operators in Eq.~(\ref{eq:intL}). The contour plot in 
Fig.~\ref{fig:contour}(a) offers a means to disentangle their coherent effects 
through a precise measurement of ${\rm BR}((\mu^+\mu^-) \to \mu^\pm e^\mp)$.
One might argue that this process alone may not constrain all operator coefficients; 
for instance, $g_{LL}^S$ could remain unconstrained at ${\rm BR} = 10^{-21}$.
However, since various experimental approaches to producing TM are under consideration 
(as discussed in the Introduction), it may be possible to collect data on TM CLFV 
decay under diverse conditions, such as using polarized muons or performing 
measurements in magnetic fields. These variations could potentially yield additional 
insight. A detailed investigation of such possibilities is deferred to future work.

Finally, we comment briefly on the experimental signature and background rejection 
for stopped TM decays. This two-body decay at rest produces a clean signal: 
$\mu^\pm$ and $e^\mp$ are emitted back-to-back with momenta 
$|\boldsymbol{p}_{\mu}| = |\boldsymbol{p}_{e}| \simeq 3m_{\mu}/4$. 
The high precision in tracking, timing, and angular resolution of charged leptons 
enables efficient signal identification using appropriate kinematic cuts.

\clearpage
\section{Summary}
\label{Sec:Summary} 

In summary, we have proposed a new process for testing CLFV and its underlying 
physics: the decay of true muonium (TM), $(\mu^+ \mu^-) \to \mu^\pm e^\mp$. 
This purely leptonic, two-body decay offers a distinct experimental signature: 
energetic, back-to-back charged leptons with a momentum of 
$|\boldsymbol{p_{\mu (e)}}| \simeq 3m_\mu/4$. 
Theoretically, it is also free from hadronic uncertainties.
Using the effective field theory approach, we evaluated the branching ratio of 
$(\mu^+ \mu^-) \to \mu^\pm e^\mp$ under the single-operator dominance scenario 
for scalar-, vector-, and dipole-type CLFV operators. Our analysis includes 
interference effects between multiple operators, which could increase the branching 
ratio. We found that the branching ratio could reach $\mathcal{O}(10^{-20})$ 
within the current experimental bounds on CLFV couplings, as constrained by the 
MEG~II experiment, for example.
Such a branching ratio may be probed at future high-luminosity muon beamlines. 
We emphasize the potential complementarity of this process with ongoing searches for new physics in rare muon decays and highlight its utility in probing the flavor structure of the lepton sector. A detailed investigation is required to firmly establish this decay as a probe of flavor dynamics. Such a study should include a comprehensive formulation of TM dynamics in both vacuum and non-vacuum environments, together with a systematic analysis of experimental backgrounds and complementarity with other CLFV processes. We leave these issues as important directions for our future work.

\section*{Acknowledgements}

The authors thank J. Sato and M. J. S. Yang for helpful discussions.
This work is supported in part by the JSPS Grant-in-Aid for
Scientific Research No.\,22K03638, 22K03602, and 20H05852 (MY). 
This work was partially supported by the MEXT Joint Usage/Research 
Center on Mathematics and Theoretical Physics JPMXP0619217849. 


%

\end{document}